%% file: fk-epl.tex
\documentstyle{europhys}
\input euromacr.tex

\begin{document}
\euro{}{}{}{}
\date{\today}

\title
{
Ground state wavefunction of the quantum Frenkel-Kontorova
model
}
\author
{
Bambi 
Hu\inst{1}\inst{2}\footnote{E-mail: bhu@hkbu.edu.hk} and
Baowen Li\inst{1}\footnote{E-mail: bwli@hkbu.edu.hk}
}
\institute{
\inst{1}Department of Physics and  Centre for Nonlinear Studies, Hong 
Kong Baptist University, Hong Kong, China \\ 
\inst{2} Department of Physics, University of Houston, Houston TX 
77204-5506, USA\\
}

\rec{ }{ }
\pacs{
\Pacs{64}{70 Rh}{ }
\Pacs{64}{70.$-$p}{ }
}
\maketitle

\begin{abstract} 
The wavefunction of an incommensurate ground state for a one-dimensional
discrete sine-Gordon model -- the Frenkel-Kontorova (FK) model --  at
zero
temperature is calculated by the quantum Monte Carlo method. It is found 
that the ground state wavefunction crosses over from an extended
state to a localized state when the coupling constant
exceeds a certain critical value.
So, although the quantum fluctuation has smeared out the breaking of
analyticity transition as observed in the classical case, the remnant of
this transition is still
discernible in the quantum regime.
\end{abstract} 

The Frenkel-Kontorova (FK) model was proposed more than half a
century
ago to 
study commensurate-incommensurate phase transitions \cite{FK38}. It has
since become a widely
used model not only in condensed matter physics but also in nonlinear
dynamics (see e.g. \cite{Selke92}). In condensed matter physics, it serves
as a model for an adsorbate layer on the surface of a
crystal \cite{Aubry,crystal}, ionic conductors and glassy
materials \cite{ion}, charge-density waves \cite{CDW}, and
chains of coupled Josephson junctions \cite{jos}. 
Moreover, the FK model is also a  nonlinear
lattice model having a normal thermal
conduction \cite{heat98}.
Recently there has been
a growing interests in applying this model to study 
dry-friction \cite{dryfriction} and atomic scale friction or 
nanotribology \cite{Klafter}. In nonlinear dynamics, the FK model
provides a vivid physical realization of the seminal 
Kolmogorov-Arnol'd-Moser theorem.

The FK model describes a chain of particles connected by  
harmonic springs and 
subject to an external sinusoidal 
potential. The classical
behavior of this model has been studied by Aubry \cite{Aubry}. As a
discrete sine-Gordon model, the FK model exhibits
many new features that are absent in the continuum sine-Gordon
model. One of the
most striking features is the so-called ``breaking of analyticity''
transition. In an
incommensurate ground state, as the coupling constant
increases beyond a critical value, the FK model displays a transition
between the unpinned state (sliding phase) and the pinned state (pinned
phase). This transition is characterized by 
the appearance of a phonon gap, the discontinuity of the
Hull function, and etc. \cite{Aubry}. These phenomena have been also found
in
many generalized FK models \cite{GFK}. 

Although a lot of studies have been done on the classical FK model, very
little study has been devoted to the quantum FK model
\cite{BAK,BGS89,BBC94,HLZ98}. Such study is also of more than academic
interest. For example, recently there has been a surge in the study
of
atomic scale friction, or nanotribology. This problem is of tremendous
importance to technology. The FK model has been used to study this
problem. However, in the nano-regime, it's inevitable that quantum
mechanics will play an important role. So an understanding of the quantum
FK model will be important to the understanding of nanotribology. 

In this Letter we would like to study the quantum version of the one
dimensional FK
model by using the quantum Monte Carlo approach. We shall
concentrate on the wavefunction of an incommensurate ground state. As we
shall see later the wavefunction changes from an extended state to
a localized one as the coupling constant is 
increased beyond a
certain critical value. This crossover is a quantum remnant
of the 
classical breaking of analyticity transition. 

The Hamiltonian operator of the one dimensional FK model is
\begin{equation}
\hat{\cal H} = \sum_i \left[\frac{\hat{p}_i^{2}}{2m} + 
\frac{\gamma}{2}(\hat{x}_{i+1} - \hat{x}_{i} - a)^2 -  
V\cos(q_0\hat{x}_i)\right]. 
\label{QMHam1}
\end{equation}
Here $m$ is the mass of the particles,
$\gamma$  the elastic constant of the spring,
$2\pi/q_0$ the period of the external potential, $V$ the magnitude
of the external potential, and $a$ the equilibrium distance between two 
nearest
neighbor particles. For convenience,
we
shall rescale the variables so as to obtain a dimensionless Hamiltonian
\begin{equation} \hat{H} = \sum_i \left[\frac{\hat{P}_{i}^{2}}{2} +
\frac{1}{2}(\hat{X}_{i+1} - \hat{X}_{i} - \mu)^2 -
K\cos(\hat{X}_i)\right], 
\label{QMHam} 
\end{equation} 
where
$K=Vq_0^2/\gamma$ is the coupling constant.
We define an ``effective Planck's constant'':
\begin{equation}
\tilde{\hbar} = \hbar\frac{q_0^2}{\sqrt{m\gamma}}.
\label{Nhbar}
\end{equation} 
This effective Planck's constant 
is the ratio of the
natural quantum energy scale $(\hbar\omega_0)$ to the natural classical
energy scale ($\gamma/q^2_0$) \cite{BBC94}, and $\omega^2_0=\gamma/m$. 

To study the ground state of Hamiltonian (\ref{QMHam}), the quantum Monte
Carlo method will be employed \cite{QMC}. As in the classical case, we
will concentrate our attention on the incommensurate state characterized
by the  golden
mean value $\sigma_G=(\sqrt{5}-1)/2$. 
In the classical case the winding number is defined as the
average separation of atoms per period, i.e.  
$\sigma=\lim_{N\rightarrow\infty}\frac{X_N-X_0}{2N\pi}$. If $\sigma$ 
is a rational 
number, the ground state is commensurate, and if it is an 
irrational number the ground state is called incommensurate.  In 
the quantum case, ${X_i} (i=0,1,\cdots, N)$ is defined as the 
expectation values of the particles.   As
usual, we use the method of continued fraction expansion and
approximate $\sigma_G$ by its rational
convergents $\sigma_G = F_n/F_{n+1}$, where $F_n$ is the $n$th Fibonacci
number
defined by the
recursion relation, $F_{n+1} = F_{n} + F_{n-1}$ with $F_0=0, F_1=1$.
Therefore, in our quantum Monte Carlo computation, we choose $F_{n+1}$
particles which are
embedded into $F_n$ external potentials with period of $2\pi$. We
impose periodic boundary condition on the chain. 

Since the external potential is periodic and has period $2\pi$, we can 
fold the wavefunction to this period and then
take the average over all particles in the interval $[0,2\pi]$. 
This quantity gives 
the probability of finding 
particles at a given potential position $X$. We plot the averaged
probability
density $\langle|\Psi|^2\rangle$ in Fig. 1 with 144 particles embedded in
89 
potentials for a fixed
$\tilde{\hbar}$ ($=0.2$) for different values of $K$.
(Notice that $\langle|\Psi|^2\rangle$ is normalized, namely, 
$\int_0^{2\pi}dX \langle|\Psi(X)|^2\rangle =1$.)
We observe that, in the  small $K$
regime, the
probability of
finding the particles at any place of the potential is almost the same
(see the curve for $K=0.1$.). This is quite similar to the sliding phase
in
the
classical version. However, as the coupling constant
increases, the
probability of finding the particle at the top of the potential decreases,
while that at the lower part of the potential increases, as is
noticeable 
from the appearance of peaks in the curves. As the potential goes
beyond a certain critical value, the probability of finding the particles
at the
top
is almost zero, see e.g. the curves for $K=2$ and $5$. 

In the classical FK model,
Coppersmith and Fisher \cite{CF83} have proposed a ``disorder parameter''
to describe the transition from the pinned phase to unpinned phase. 
This parameter is defined as the minimum distance of a particle from
the top of a well, $ {\it D}_{cl}
=\min_{j,n}|X_j^{cl} - 2\pi(n+\frac{1}{2})|$. It can be seen
that, as long as $D_{cl} \neq 0$, the particles
are pinned, while
$D_{cl}=0$ corresponds to an unpinned state.
This ``disorder parameter'' also measures the discontinuity (or width of
the biggest gap) of the Hull function.  
One might naively try to use the same function to describe the
quantum crossover. For instance, one may 
define a very similar quantum disorder parameter
$D_q=\min_{j,n}|X_j^q-2\pi(n+\frac{1}{2})|$, where $X_j^q$ is the 
expectation value of the $j$'th particle's position. However, this parameter 
$D_q$ could not capture the crossover of the ground state wavefunction.
The 
reason is that, in the quantum case, the particle can tunnel from one
side of 
the potential to the other, thus the gap in the classical Hull function
does 
not survive the quantum fluctuations (see Refs.
\cite{BGS89} and  \cite{HLZ98}). It turns out that even if $K>K_c^q$, 
$D_q$ might be also close to (or equal to) zero.
Therefore, 
a new parameter is needed to describe the wavefunction crossover in the
quantum FK 
model. To
this end, we define the probability of finding particles at the potential
top as a quantum ``disorder parameter'', which is denoted as $P_t$. $P_t$
is
given by
\begin{equation}
P_t =\frac{1}{N}\sum_{n=0}^{N-1} \int 
|\Psi(X)|^2\delta\left(X-2\pi(n+\frac{1}{2})\right)dX. 
\label{Pt}
\end{equation}

In Fig. 2(a)-(b) we plot $P_t$ versus $K$ for different 
temperature and different system size. In Fig. 2(a), 
we fix the quantum fluctuation $\tilde{\hbar}(=0.2)$, and the
particle numbers (13/21) by changing the temperature from
$T_e=0.2/30,
0.2/120,$ to
$0.2/480$. The convergence is quite fast as is seen from the figure. The
sharp decrease of $P_t$ is very evident for $K$ between 1 and 2. In this
small $K$ regime, $P_t$ changes dramatically: it drops almost two
orders of magnitude. This dramatical change can also be seen from Fig.2
(b). There the temperature is fixed at
$T_e=\tilde{\hbar}/N\delta\tau=0.2/120$, which can  be regarded as
effectively zero
temperature.  The system size is changed from 21 particles to 144
particles. In the quantum Monte Carlo method 
the Feynman path integral presentation is used. In this presentation,
solving a one dimensional quantum problem is equivalent to solving a two
dimensional effective
classical problem with an extra dimension as the imaginary
time \cite{RMP97}. The size of this extra dimension is
$\tilde{\hbar}\beta$, where $\beta=1/T_e$, $T_e$ is the temperature.
Therefore, in the zero temperature limit, we shall take $\beta$ infinite. 

It is worth pointing out that
another parameter which can be used to depict this crossover is the
maximal fluctuations of the particles. In the 
quantum case, we have observed taht the particle situated at a position
nearest to the maxim of the potential  has always
maximal
fluctuation since it has the largest classical potential energy. This
observation  has been verified numerically by the quantum Monte Carlo
results, and
theoretically by
our squeezed state theory \cite{HLZ98}.  Thus, we can use this maximal
fluctuation as another measure of quantum ``disorder''. To make a 
qualitative comparison with
the  classical disorder parameter $D_{cl}$ defined by Coppersmith and 
Fisher \cite{CF83}, we 
shall make use of the standard deviation, i.e. the square root of the
maximal fluctuation, 
\begin{equation}
\Delta = \max_j \left\{\left(\langle X_j^2\rangle - \langle 
X_j\rangle^2\right)^{\frac{1}{2}}\right\}.
\label{DQM}
\end{equation}
Here the average $\langle\cdots\rangle$ is taken over all the paths 
produced in the quantum Monte Carlo simulation, which is about 4,000 in
our calculations.
$\Delta$ versus $K$ is plotted in
Fig. 3. The computations given in this figure have been
carried out with $\sigma=F_n/F_{n+1}=34/55$. 

It is noticeable that the transition of the wavefunction is characterized
by the different $K$-dependent behavior of $\Delta$.  For $K<K_q^*$,
$\Delta$ is a constant. It does not depend on $K$, but it
changes with $\tilde{\hbar}$. For $K>K_q^*$, the maximal fluctuation
increases with $K$, but it does not change with $\tilde{\hbar}$.
Furthermore, this
quantum parameter $\Delta$ is approximately equal to the classical
disorder
parameter $D_{cl}$. For comparison with the classical disorder parameter,
we also include $D_{cl}$  in the inset of Fig. 3. In the classical case,
the transition occurs at $K=K_c^*$, where  $K_c^*= 0.971635...$. 

These results show that the width and the shape of the probability density
do not
change by changing the external potential strength in the small $K
(<K_q^*)$
regime. The width only depends on the strength of the quantum fluctuation
$\tilde{\hbar}$. However, this picture changes dramatically when
$K >K_q^*$. In this regime, the profile of
the
probability density spreads out, and the width of the probability density
is insensitive to quantum fluctuations; instead it depends only on the
coupling constant. 
It must be stressed that the analogy between $D_{cl}$
and $\Delta$ cannot carried too far, since $\Delta$ is not exactly the 
quantum correspondence of $D_{cl}$. This is why even if we let
$\tilde{\hbar}$ go to zero, $\Delta$ would not be zero in the regime of $K
< K_q^*$.

In summary, we have studied the ground state wavefunction of the FK
model at zero temperature. The wavefunction undergoes a crossover
from an extended state (analogous to the sliding phase) to a localized
state (analogous to the pinned phase) as the external potential increases.
Therefore,
although the quantum fluctuation has smeared out the breaking of
analyticity transition as observed in the classical case, the remnant of
this transition is still
discernible in the quantum regime.

\stars
We would like to thank S. Aubry, F. Borgonovi, D. K. Campbell, H. 
Chen,
R. B.  Griffiths, H.-Q. Lin, L.-H. Tang, and W.-M. Zhang for
helpful discussions during several periods of the work. BL thanks the
Abdus Salam International Centre for Theoretical Physics (Trieste) for its
kind hospitality during his visit there in the summer of 1998.
This work was supported
in part by the grants from the Hong Kong Research Grants Council (RGC) and
the Hong Kong Baptist University Faculty Research Grant (FRG).

\section{Figures}
\begin{figure}
\caption{The wavefunction of an incommensurate ground state  having
winding number 89/144 at fixed $\tilde{\hbar}=0.2$ for different values of 
$K$. The
wavefunction $\langle|\Psi|^2\rangle$ means the probability of finding the 
particles at
$X$ (mod $2\pi$). The curves are for $K=0.1, 1, 2$, and $5$, respectively.
The wavefunction becomes localized at the lower part of the potential as
$K$ changes to 2 and 5. 
}
\end{figure}

\begin{figure}
\caption{The quantum ``disorder parameter'' $P_t$, i.e. the probability of
finding
particles on the top of the external potential versus $K$ for a fixed  quantum
fluctuation $\tilde{\hbar}=0.2$. 
(a) $P_t$ for different temperature at a fixed winding number
13/21. The temperature changes from $0.2/30, 0.2/120$, and $0.2/480$,
respectively.
(b) $P_t$ for different winding numbers: 13/21, 34/55, and 89/144, at fixed
temperature $T=0.2/120$, which can be effectively regarded as zero. 
}
\end{figure}

\begin{figure} 
\caption{ $\Delta$ versus $K$. Different symbols represent different
effective Planck's
constant, which are $\tilde{\hbar}= 0.01$, and 0.1, respectively. We
draw the classical disorder parameter defined by Coppersmith and
Fisher [17] in the inset for comparison. The winding number  for all 
calculations given in this figure is 34/55.  
} 
\end{figure}


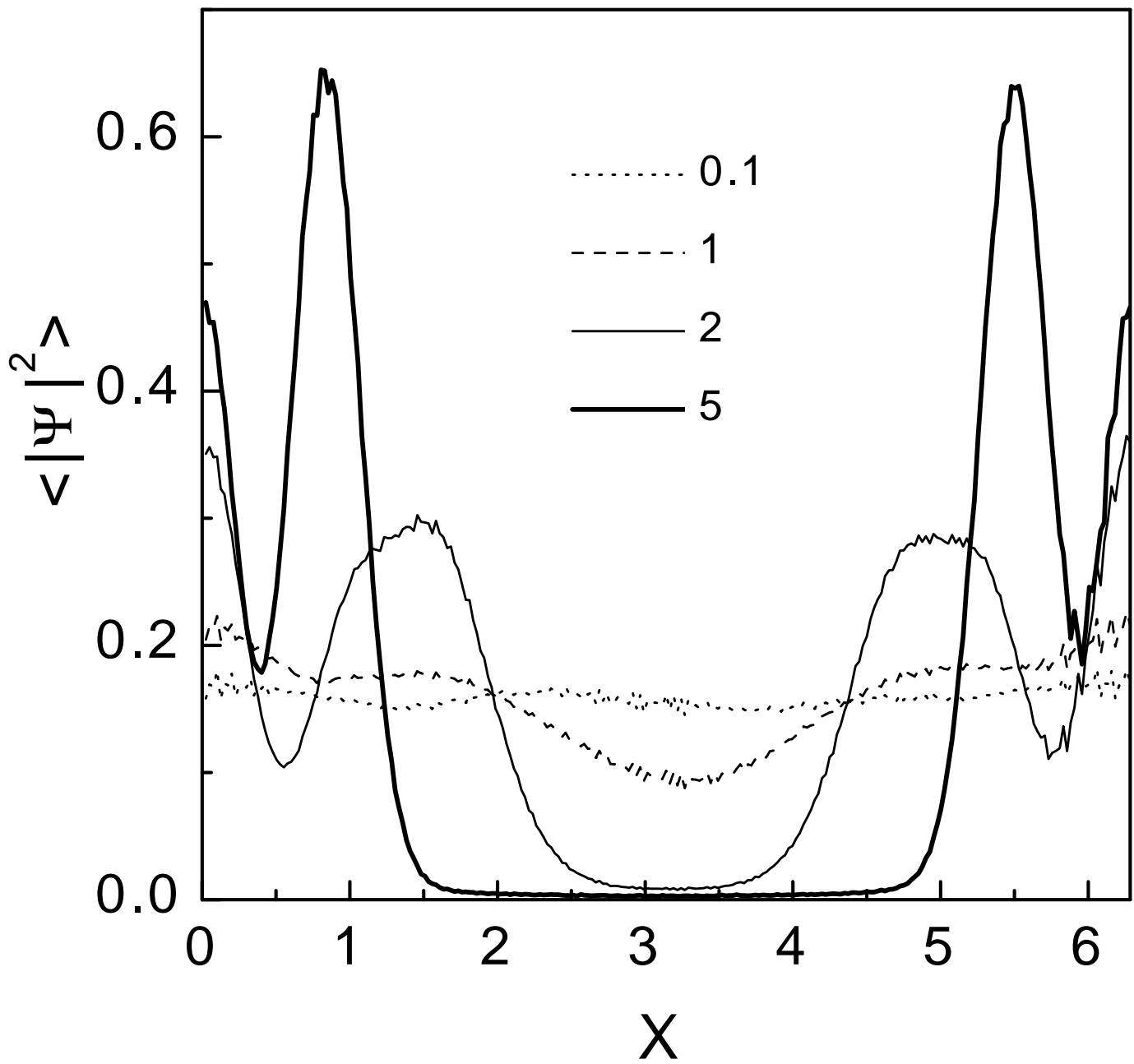
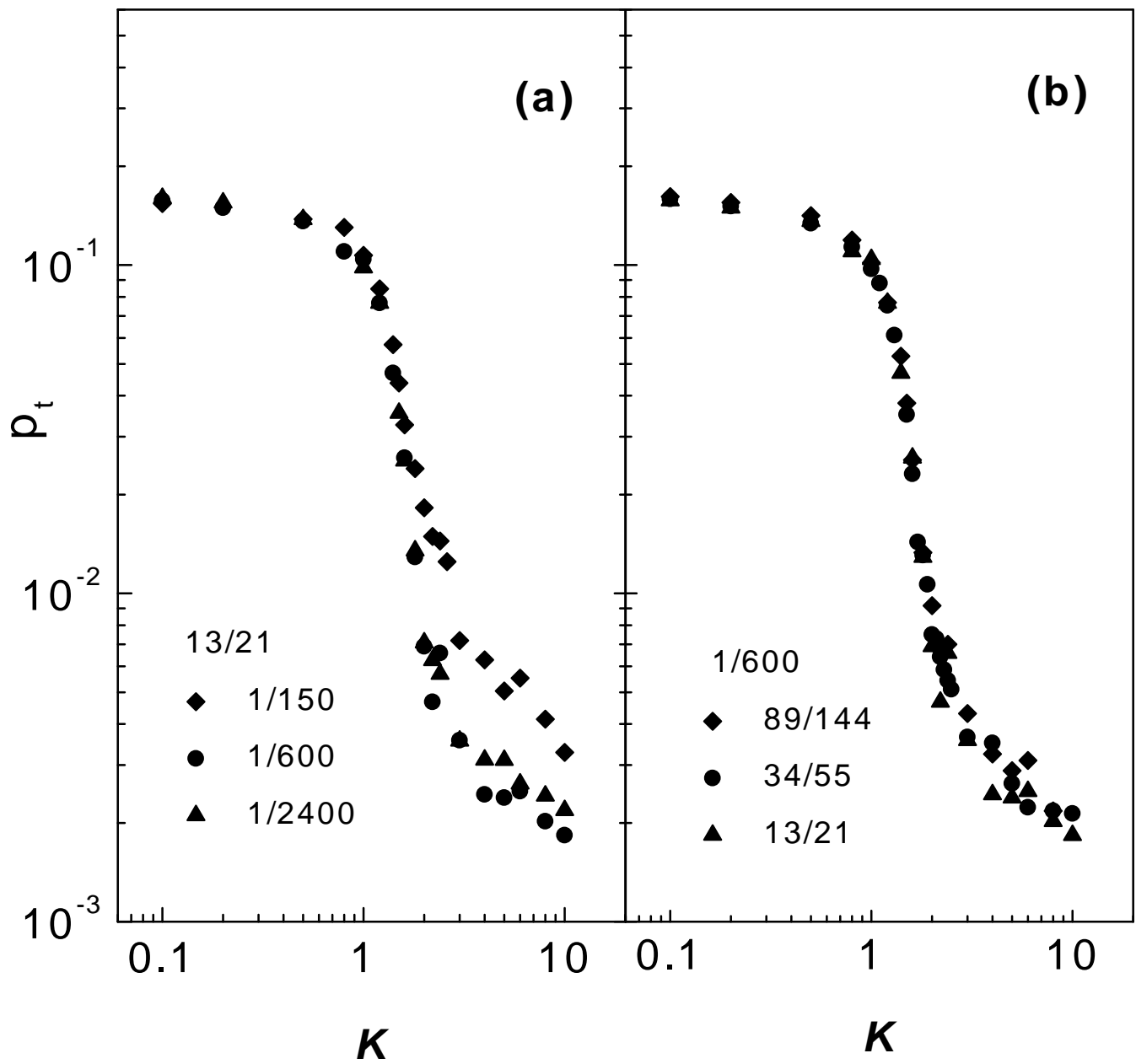
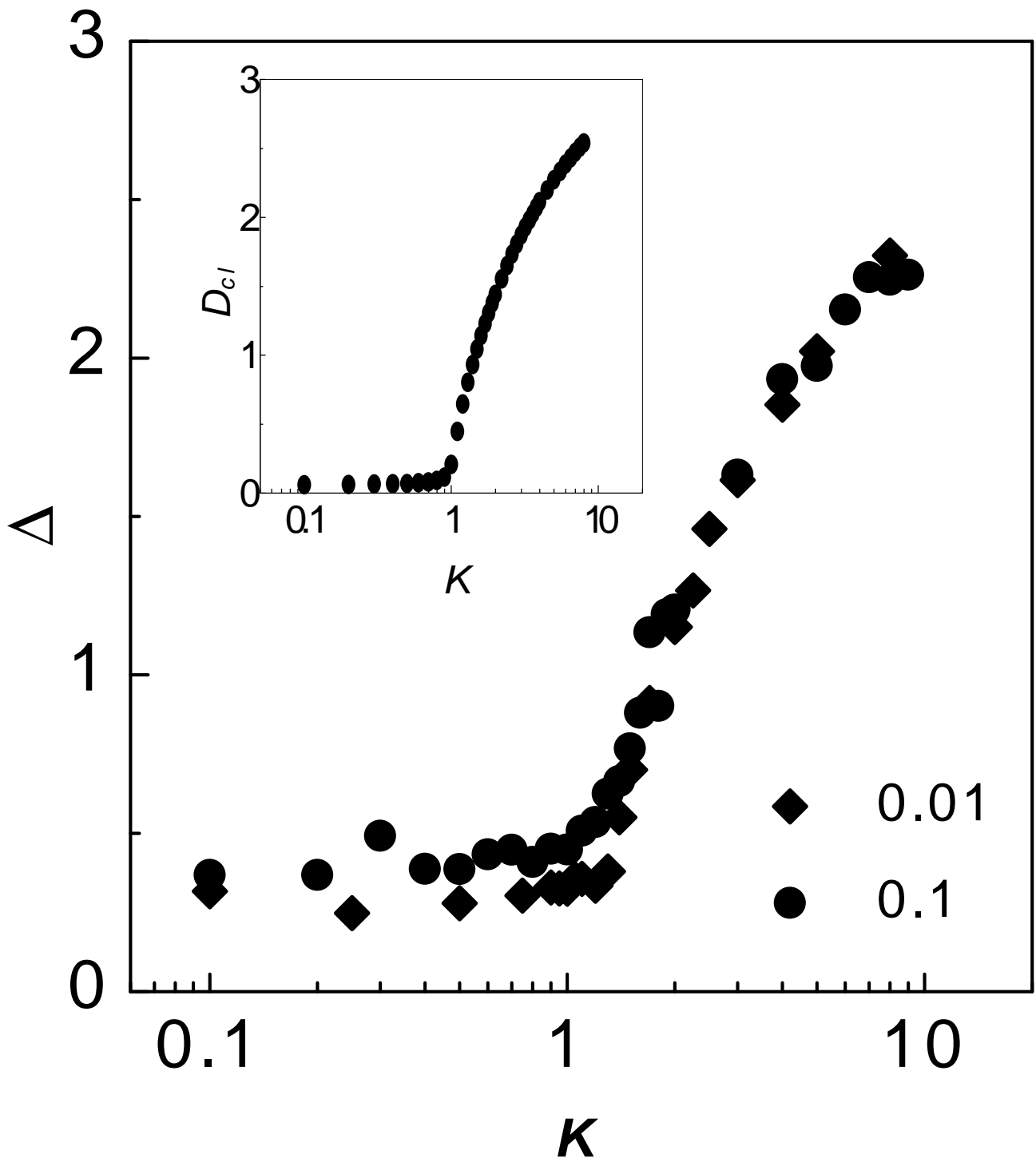
\end{document}

%% file: euromacr.tex
\def\stars{\bigskip\centerline{***}\medskip}

\newif\ifboo \boofalse

\def\Review#1{\boofalse{\it #1},}
\def\Name#1{{\sc #1},}
\def\Vol#1{\ifboo Vol. {\bf #1}\else{\bf #1}\fi}
\def\Year#1{\ifboo #1\else(#1)\fi}
\def\Book#1{\bootrue{\it #1},}
\def\Page#1{\ifboo {\rm p. #1}\else{\rm #1}\fi}

%% file: fk-epl.bbl
\begin{thebibliography}{99}

\bibitem{FK38}
\Name{Frenkel Y and Kontorova T K} 
\Review{Zh.~Eksp.~Teor.~Fiz.}
\Vol{8} 
\Year{1938} 
\Page{1340};
\Name{Frank F and van der Merwe J}
\Review{Proc. R. Soc. London A }
\Vol{198}, 205 
\Year{1949}
\Page{205}.

\bibitem{Selke92}
\Name{Bak P}
\Review{Rep.Prog. Phys.}
\Vol{45}
\Year{1982}
\Page{587} and 
references therein; 
\Name{Selke W} in 
\Book{Phase Transition and Critical 
Phenomena} 
ed. \Name {Domb C and Lebowith J L},
\Vol{15}. Academic Press \Year{1992}, London.

\bibitem{Aubry}
\Name{Aubry S} in \Book{Solitons and Condensed Matter Physics}, 
edited by \Name{Bishop A. R. and Schneider T}. (Springer, New York, 
\Year{1978}; 
\Name{Aubry S}
\Review{J. Phys. (France)} 
\Vol{44}
\Year{1983}
\Page{147}; 
\Name{Peyrard M and Aubry S} 
\Review{J. Phys. C}
\Vol{16}
\Year{1983}
\Page{1593}; 
\Name{Aubry S}
\Review{Physica D}
\Vol{7}
\Year{1983}
\Page{240};
\Name{Aubry S and Le Da\"eron P Y} {\it ibid}
\Vol{8}
\Year{1983}
\Page{381}. 

\bibitem{crystal} 
\Name{Uhler W and Shilling R}
\Review{Phys. Rev. B} \Vol{37} \Year{1988} \Page{5758}.

\bibitem{ion} 
\Name{Pietronero L, Schneider W R, and Str\"assler S}
\Review{Phys. Rev. B} \Vol{24} \Year{1981} \Page{2187}; 
\Name{Aubry S}
\Review{J. Phys. (France)} 
\Vol{44} 
\Year{1983} 
\Page{147};
\Name{Vallet F, Schilling R, and Aubry S} \Review{J. Phys. C} \Vol{21}
\Year{1988} \Page{67}.

\bibitem{CDW}
\Name{Flor\'ia L M and Mazo J J}
\Review{Adv. Phys.} 
\Vol{45} 
\Year{1996} 
\Page{505}.

\bibitem{jos}
\Name{Watanabe S, van der Zant H S J, Strogatz S H , and Orlando T P} 
\Review{Physica D} 
\Vol{97} 
\Year{1996} 
\Page{429}.


\bibitem{heat98}
\Name{Hu B, Li B, and Zhao H}
\Review{Phys. Rev. E}
\Vol{57}
\Year{1998}
\Page{2992} ;
\Name{Fillipov A, Hu B, Li B, and Zeltser A}
\Review{J. Phys. A}
\Vol{31}
\Year{1998}
\Page{7719} ;
\Name{Tong P.-Q, Li B, and Hu B}
\Review{Phys. Rev. B}
\Vol{59}
\Year{1999}
\Page{8639}.

\bibitem{dryfriction}
\Name{Braun O M, Dauxois T, Paliy  M V, and Peyrard M}\Review{Phys. Rev. 
Lett.} \Vol{78}
\Year{1997} \Page{1295}; 
\Review{Phys. Rev. E} 
\Vol{55} 
\Year{1997} 
\Page{3598}; 
\Name{Braun O M, Bishop A R, and R\"oder J} 
\Review{Phys. Rev. Lett.} 
\Vol{79} 
\Year{1997} 
\Page{3692}.
\Name{Weisee M and Elmer F J} 
\Review{Phys. Rev. B} 
\Vol{53} 
\Year{1996} 
\Page{7539};
\Name{Stunz T and Elmer F J} 
\Review{Phys. Rev. E}
\Vol{58} 
\Year{1998} 
\Page{1602, 1612};
\Name{Zheng Z G, Hu B, and Hu G} \Review{Phys. Rev. B} \Vol{58}
\Year{1998} \Page{5453}.

\bibitem{Klafter}
\Name{Rozman M G, Urbakh M, and Klafter J}
\Review{Phys. Rev. Lett.}
\Vol{77}
\Year{1996}
\Page{683};
\Name{Zaloj V, Urbakh M, and Klafter J}
\Review{Phys. Rev. Lett.}
\Vol{81}
\Year{1998}
\Page{1227}.

\bibitem{GFK}
\Name{Johannesson H, Schaub B and  Suhl H} \Review{Phys. Rev. B} 
\Vol{37} 
\Year{1988} 
\Page{9625}; 
\Name{Lin B and Hu B} \Review{J. Stat. Phys.} 
\Vol{69} 
\Year{1992}
\Page{1047};
\Name{Shi J and Hu B} \Review{Phys. Rev. A} 
\Vol{45} 
\Year{1992} 
\Page{5455}; 
\Name{Chou C I, Ho C L, and Hu B}
\Review{Phys. Rev. E} \Vol{55} \Year{1997} \Page{5092}; \Name{Chou C I, 
Ho C L, Hu B, and Lee H} {\it ibid.} \Vol{57} \Year{1998} \Page{2747}; 
\Name{Xu A G, Wang G R, Chen S G, and Hu B} \Review{Phys. Rev. B} \Vol{58}
\Year{1998} \Page{721};
\Name{Xu A G, Wang G R, Chen S G, and Hu B} \Review{Phys. Rev. B}
\Vol{57} \Year{1998} \Page{2771}.


\bibitem{BAK}
\Name{Bak P and Fukuyama H} \Review{Phys. Rev. 
B} \Vol{21} \Year{1980} \Page{3287}.

\bibitem{BGS89}
\Name{Borgonovi F, Guarneri I and Shepelyansky D} \Review{Phys. Rev. 
Lett.} \Vol{63} \Year{1989} \Page{2010}; \Review{Z. Phys. B} 
\Vol{79} \Year{1990} 
\Page{133};
\Name{Borgonovi F}, \Book{Ph. D dissertation}, Universit\'a Degli 
Studi di Pavia, \Year{1989}, Italia.

\bibitem{BBC94}
\Name{Berman G P, Bulgakov E N and Campbell D K} \Review{Phys. Rev.
B} \Vol{49} \Year{1994} \Page{8212}.

\bibitem{HLZ98}
\Name{Hu B, Li B, Zhang W M} \Review{Phys. Rev. 
E} \Vol{58} \Year{1998} \Page{R4068}.

\bibitem{QMC}
\Name{Creutz M and Freedman B} \Review{Ann. 
Phys.} \Vol{132} \Year{1981} \Page{472}; 
\Name{Shuryak E V and  Zhirov O V} \Review{Nucl. Phys. 
B} \Vol{242} \Year{1984} \Page{393}.

\bibitem{RMP97}
\Name{Sondhi S L, Girvin S M, Carini J P, and Shahar D} \Review{Rev. Mod. 
Phys.} \Vol{69} \Year{1997} \Page{315}.

\bibitem{CF83}
\Name{Coppersmith N and Fisher D S} \Review{Phys. Rev. 
B} \Vol{28} \Year{1983} \Page{2566}.

\end{thebibliography}
